\documentclass{aip-cp}

\usepackage[numbers]{natbib}
\usepackage{adjustbox}
\usepackage{multicol}
\usepackage{hhline}
\usepackage{rotating}
\usepackage{amstext}
\usepackage{graphicx}
\usepackage[implicit=false]{hyperref}
\usepackage{xcolor}

\hypersetup{
    colorlinks,
    linkcolor={red!50!black},
    citecolor={blue!50!black},
    urlcolor={blue!80!black}
}

\begin{document}

\title{Survey of deformation in nuclei in order to estimate the enhancement of sensitivity to atomic EDM}

\author[aff1]{Prajwal Mohanmurthy \corref{cor1}}
\author[aff2]{Umesh Silwal \noteref{note2}}
\author[aff2]{Durga P. Siwakoti}
\author[aff2]{Jeff A. Winger \corref{cor2}}

\affil[aff1]{Laboratory for Nuclear Science, Massachusetts Institute of Technology, Cambridge, MA 02139}
\affil[aff2]{Department of Physics and Astronomy, Mississippi State University, Mississippi State, MS 39762-5167}
\corresp[cor1]{Corresponding author, electronic address: \href{mailto:prajwal@mohanmurthy.com}{prajwal@mohanmurthy.com}}
\corresp[cor2]{Corresponding author, electronic address: \href{mailto:j.a.winger@msstate.edu}{j.a.winger@msstate.edu}}
\authornote[note2]{Speaker; currently at Department of Physics and Astronomy, University of Wyoming, Laramie, WY 82071.}

\maketitle

\begin{abstract}
The observed baryon asymmetry of the universe (BAU) cannot be explained by the known sources of charge-parity (CP)-violation in the Standard Model (SM). A non-zero permanent electric-dipole-moment (EDM) for fundamental particles, nuclei or atoms, violates CP. Measuring a non-zero EDM allows us to gain a handle on additional sources of CP-violation required to explain the observed BAU. The EDM of an atom with an octupole and quadrupole deformed nucleus is enhanced. Therefore, the search for such atoms has become important in the quest to measure an EDM. Viability of EDM searches in $^{225}$Ra atoms with a deformed nucleus have already been demonstrated. We have performed a comprehensive survey for possible EDM candidates from a list of octupole deformed nuclei predicted by various theoretical models. Our search of long-lived isotopes with nuclear deformations comparable to or better than $^{225}$Ra results in a handful of viable candidates for future atomic EDM experiments based out of the Facility for Rare Isotope Beams (FRIB): $^{221}$Rn, $^{221,223,227}$Fr, $^{221,223,225}$Ra, $^{223,225,227}$Ac, $^{229}$Th, and particularly $^{229}$Pa. Furthermore, nuclei of $^{223,225}$Rn, $^{225}$Fr and $^{226}$Ac are also highly quadrupole and octupole deformed, but their ground state parity doublet energy difference has not yet been measured.
\end{abstract}

\vspace{-0.25cm}
\section{INTRODUCTION}

Lorentz invariance is an integral part of the Standard Model (SM), and the SM has been shown to be invariant under the joint transformations of charge conjugation (C) - parity (P) - time reversal (T): \emph{i.e.} SM is CPT invariant \cite{[3-1],[3-2]}. Furthermore, the CPT theorem demonstrated that if CPT symmetry is violated, Lorentz symmetry is also violated \cite{[3-3]}. This firmly places the SM in a CPT and Lorentz invariant domain. Even though CPT is conserved, sub-symmetries such as CP, P, or T, to name a few, can be violated.

Most of the observable universe is made of matter as opposed to equal amounts of matter and anti-matter. Particularly, baryon/anti-baryon asymmetry is critical to nucleosynthesis in the early universe \cite{[1]}. The baryon asymmetry of the universe (BAU) has been very precisely measured with the help of cosmic microwave background observations, the latest value being reported by the Planck telescope \cite{[2]}. Andrei Sakharov identified the minimum requirements for interactions to generate a BAU, \emph{viz.} (i) baryon number violation, (ii) violation of both charge conjugation symmetry and the joint charge CP symmetry, and (iii) interactions out of the thermal equilibrium \cite{[3]}. No baryon number violating process has been measured yet, where experiments in search of neutron/anti-neutron oscillation \cite{[5-1]}, neutron/mirror neutron oscillation \cite{[5-2-1],[5-2-2],[5-2-3]}, and proton decay \cite{[5-3]} have set very stringent limits on baryon number violation. The last Sakharov condition: (i) ensures that CPT symmetry does not force baryon number conservation \cite{[5-4]}, and (ii) sets the rate of baryon number violating interactions below the expansion rate of the universe \cite{[5-5]}.

In this paper we are mostly concerned with the second Sakharov condition. CP violation has been observed in weak interaction mediated processes such as kaon \cite{[4]} and B-meson \cite{[5]} decays. Thus far, no CP violation has been observed in electromagnetism or quantum chromodynamics (QCD) mediated interactions. After all, in the quark sector of the SM, CP violation is only present in the quark mixing matrix, \emph{i.e.} the Cabbibo-Kobayashi-Maskawa (CKM) matrix; and the CKM matrix elements quantify the weak interaction in the quark sector. But the amount of CP violation in the CKM matrix is insufficient to explain the observed BAU \cite{[5-5]}. Therefore we need new sources of CP violation beyond what is in the CKM matrix. It is rather straight forward to introduce CP violation into QCD, through the QCD-$\theta$ term \cite{[5-6]}. But other extensions to the SM, such as supersymmetry, could also add to the amount of CP violation \cite{[5-7]}. Measuring a permanent EDM in multiple systems provides for a lucrative means to understand the sources of CP violation in the universe. Efforts to measure permanent CP violating EDMs for subatomic particles, atoms, and molecules have been underway \cite{[14]}, and feature prominently in the US long range plan \cite{[15]}.

\vspace{-0.25cm}
\section{MULTIPOLE EXPANSION OF THE NUCLEAR POTENTIALS}

Electromagnetic scalar and vector potentials for a distribution of charges in motion may be written in terms of a multipole expansion. For the purposes of this paper, we are concerned with the description of the nuclear electric and magnetic content in terms of a multipole expansion. From the Purcell-Ramsey-Schiff theorem \cite{[6-0-1],[16]}, the scalar potential, $\phi({\bf R})$, may be written as \cite{[23]}:
\begin{eqnarray}
\phi({\bf R})&=&\underbrace{\int d^3{\bf r} \cdot \frac{\rho({\bf r})}{|{\bf R}-{\bf r}|}}_{\phi^{(0)}({\bf R})} + \underbrace{\frac{1}{Ze}{\bf d}\int d^3{\bf r} \cdot \nabla \frac{\rho({\bf r})}{|{\bf R}-{\bf r}|}}_{\phi^{(1)}({\bf R})} + \mathcal{O}\left(\phi^{(2)}({\bf R})\right)\label{eq1},
\end{eqnarray}
where $\rho({\bf r})$ is the nuclear charge density such that $\int d^3{\bf r}\cdot \rho({\bf r}) = Ze$, $Z$ is the number of protons in the nucleus, and ${\bf d} = \int d^3{\bf r}\cdot {\bf r}\rho({\bf r})$ is the electric dipole moment (EDM). Individual terms of the above multipole expansion, $\phi^{(l)}({\bf R})$, can be written in terms of electric $2^l$-pole moments, $Q^{(e)}_{lm}$, and spherical harmonics, $Y_{lm}(\Theta,\Phi)$, \cite{[6-0-2]}:
\begin{eqnarray}
\phi^{(l)}({\bf R}) &=& \frac{1}{R^{l+1}}\sum_{m=-l}^{m=l}\sqrt{\frac{4\pi}{2l+1}}Q^{(e)}_{lm}Y_{lm}(\Theta,\Phi)\label{eq2}, \\
\text{where,}\quad~Q^{(e)}_{lm} &=& \sqrt{\frac{4\pi}{2l+1}}\int d^3r\cdot r^l \rho({\bf r})Y_{lm}\left(\frac{{\bf r}}{r}\right)\label{eq3}.
\end{eqnarray}
In this description of the nuclear scalar potential, the term corresponding to $l\!=\!0$ would correspond to an electric mono-pole moment, $eZ$; the $l\!=\!1$ term would correspond to the electric dipole moment, ${\bf d}=\left<|Q^{(e)}_{(l=1)(m=-1)}|,|Q^{(e)}_{(l=1)(m=0)}|,|Q^{(e)}_{(l=1)(m=1)}|\right>$; and so forth.

Similarly, the vector potential of the nucleus, ${\bf A}({\bf R})$, can be written in terms of a multipole expansion, and the individual terms of the expansion may in turn be written in terms of magnetic $2^l$-pole moments, $Q^{(m)}_{lm}$:
\begin{eqnarray}
Q^{(m)}_{lm} &=& \frac{1}{l+1}\sqrt{\frac{4\pi}{2l+1}}\int d^3r\cdot ({\bf r} \times {\bf j}) \nabla\left[r^l \rho({\bf r})Y_{lm}\left(\frac{{\bf r}}{r}\right)\right]\label{eq4},
\end{eqnarray}
where, ${\bf j}={\bf l}+{\bf s}$, is the total angular momentum associated with the nucleus. Like in the case of electric $2^l$-pole moments, here, the term associated with $l\!=\!0$ would correspond to a magnetic mono-pole moment ($Q^{(m)}_{00}=0$); the $l\!=\!1$ term would correspond to the magnetic dipole moment (MDM); and so forth.

The Wigner-Eckart theorem \cite{[6]} suggests that a system with spin $s$ can possess up to and including $2^{2s}$-pole electromagnetic moments. So, a $0^+$ nucleus only possesses an electric charge, and is spherically symmetric. It follows from this that a system needs to be at least spin $1/2$ to describe dipole moments (EDM and MDM); at least spin $1$ to describe quadrupole moments (EQM and MQM); and so forth. Electric $2^{2\mathcal{N}}$-pole moments and magnetic $2^{2\mathcal{N}+1}$-pole moments are T and P even, where $\mathcal{N}$ is a whole number, while electric $2^{2\mathcal{N}+1}$-pole moments and magnetic $2^{2\mathcal{N}}$-pole moments are T and P odd.  Hence, we need a system with at least a non-zero EDM and a non-zero MDM in order for the system to violate P and T symmetry. Since CPT is conserved in the SM, such a system would also violate CP and CT symmetries. In fact, a non-zero EDM in the ground state of any sub-atomic particle (like protons, neutrons, or electrons) or an atom (with ${\bf j}>0$) violates P and T symmetries \cite{[7]}.

\vspace{-0.25cm}
\section{PERMANENT ELECTRIC DIPOLE MOMENT OF ATOMS}

In the SM, the CKM-matrix along with the QCD-$\theta$ parameter endow sub-atomic particles with T, P, CT and CP violating EDMs, which in turn contribute to a non-zero EDM of the atom. The neutron (and proton) acquires an EDM of $d^{\text{(CKM)}}_n\sim2\times10^{-32}\,e\cdot cm$ from the CKM matrix \cite{[8]}, while the EDM from QCD-$\theta$ term is $d^{\text{QCD-}\theta}_n\sim\theta(6\times10^{-17})\,e\cdot cm$ \cite{[9]}. Experiments have placed an upper limit on the value of neutron EDM, $d_n < 3\times10^{-26}\,e\cdot cm$ (90 \% C.L.) \cite{[10]}, which constrains the value of $\theta<5\times10^{-10}$. The origins of the smallness for QCD-$\theta$ is unknown and is commonly referred to as the strong-CP problem. So, in the case of the neutron, any non zero EDM can be explained within the SM, and needs no contribution from extensions to the SM. On the other hand, the CKM EDM of the electron is $d^{\text{(CKM)}}_e\sim10^{-44}\,e\cdot cm$ \cite{[11]}, and its QCD-$\theta$ EDM is $d^{\text{QCD-}\theta}_e\sim-\theta(2.2 \textrm{--} 8.6)\times10^{-28}\,e\cdot cm$ \cite{[12]}. Using the constraint on QCD-$\theta$, the electron EDM from the QCD-$\theta$ is constrained to $d^{\text{QCD-}\theta}_e<-(1.1 \textrm{--} 4.3)\times10^{-37}\,e\cdot cm$, but experiments only constrain the electron EDM to $d_e<1.1\times10^{-29}\,e\cdot cm$ \cite{[12-2]}. In the case of an electron, an EDM larger than   $4.3\times10^{-37}\,e\cdot cm$ would require contributions from extensions to the SM.

Classically, the electron cloud of an atom would screen the nuclear EDM \cite{[16]}. However, Schiff screening of the nuclear EDM by the surrounding electron cloud may not be perfect \cite{[17]} if, (i) the electrons are relativistic, as is the case with paramagnetic atoms with a single unpaired valence electron, such as $^{85}$Rb \cite{[18-1]}, $^{133}$Cs \cite{[18-2]}, $^{205}$Tl \cite{[18-3]}, and $^{210}$Fr \cite{[18-4]}, (ii) the nucleus has quadrupole and octupole deformations, as is the case with diamagnetic atoms of $^{223}$Rn \cite{[19-1]} and $^{225}$Ra \cite{[19-2]}, or if (iii) there exists dominant CP violating interactions between the constituents of the atom, as is the case with the diamagnetic atoms of $^{129}$Xe \cite{[20-1]} and $^{199}$Hg \cite{[20-2]}. Generally, paramagnetic atoms acquire an EDM through amplification of the electron EDM, whereas, diamagnetic atoms acquire an EDM through improper screening of the nuclear EDM. The residual of the improper screening of the nucleus is called the Schiff moment.

Considering only the contribution from the CKM matrix and the maximum allowed contribution from the QCD-$\theta$ term, the EDM of individual nucleons, such as protons and neutrons, is larger than that of individual leptons, such as electrons, by over $12$ orders of magnitude. For diamagnetic atoms, such as $^{129}$Xe and $^{199}$Hg, the expectation of atomic EDM in the CKM+QCD-$\theta$ framework, is small due to Schiff screening. Schiff screening in such diamagnetic atoms leads to an atomic EDM which is usually smaller than that of individual nucleons by $\sim(3\textrm{--}5)$ orders of magnitude \cite{[21]}. In the case of paramagnetic atoms, where the electron EDM is enhanced by some $\sim(1\textrm{--}3)$ orders of magnitude \cite{[21]}, the expectation of atomic EDM, in the CKM+QCD-$\theta$ framework, is still extremely small due to the smallness of the electron EDM, even smaller than that of diamagnetic atoms.

\vspace{-0.25cm}
\section{ENHANCEMENT OF ATOMIC ELECTRIC DIPOLE MOMENT DUE TO NUCLEAR DEFORMATIONS}

The diamagnetic atoms, such as $^{223}$Rn and $^{225}$Ra, that have deformed nuclei, present an interesting class of systems. The nuclear deformation enhances the atomic EDM (of $^{223}$Rn and $^{225}$Ra) when compared to other diamagnetic atoms (such as $^{129}$Xe and $^{199}$Hg). For example, the nucleus in both $^{225}$Ra and $^{223}$Rn are comparably deformed, and their atomic EDM is expected to be enhanced by around $\sim(2\textrm{--}3)$ orders of magnitude \emph{w.r.t.} $^{199}$Hg \cite{[22],[23]}. In this section, we will take a look at the parameters that determine this enhancement factor.

Nuclear structure deformations are also characterized with a multipole description, similar to the nuclear charge content. The even-even core of the nucleus contributes significantly to the charge description of the nucleus \cite{[23],[24]}. The axially symmetric core is sufficiently described by the two-fluid liquid drop model \cite{[25-1],[25-2],[25-3],[25-4]}, the surface of which can be written as:
\begin{eqnarray}
R=c_V R_0 \left(1+\sum_{l=1}\beta_l Y_{l0}\right)\label{eq5},
\end{eqnarray}
where $c_V=1-\{(1/\sqrt{4\pi})\sum_{l=1}\beta^2_l\}$ is the normalization, $R_0=1.2~\text{fm}\cdot A^{1/3}$, $A$ is the total number of nucleons in the nucleus, and $\beta_l$ are the $2^l$-pole structure deformation coefficients. It is important not to mix the multipole description of the charge content of the nucleus with nuclear deformations. For example, while a $j=1/2$ nucleus can only posses up to and including electric and magnetic dipole moments, such nuclei could still be structurally octupole deformed.

The Schiff moment in an atom with non-zero quadupole and octupole deformations is given by \cite{[23],[24],[26],[27]}:
\begin{eqnarray}
S &\approx& \beta_2\beta^2_3 Z A^{2/3}\left(\frac{1}{E_- - E_+}\right) \frac{j}{j+1}\eta ~\cdot~ 10^{-4}\,\text{e fm}^3,\label{eq6}
\end{eqnarray}
where $\eta$ is the strength of the P, T violating potential of the nucleus, and $E_{\pm}$ are the energies of the ground state parity doublet in keV. Atomic EDM is proportional to the Schiff moment \cite{[24],[28]}:
\begin{eqnarray}
d_{\text{atom}} \approx SZ^2R_{1/2}~~~ \Rightarrow ~~~d_{\text{atom}} \propto \frac{\beta_2\beta^2_3 Z^3 A^{2/3}}{E_- - E_+}\label{eq8},
\end{eqnarray}
where $R_{1/2} = (4\gamma_{1/2}/\{\Gamma(2\gamma_{1/2}+1)\}^2)(2ZR_0/a_B)^{2(\gamma_{1/2}-1)}$ is the relativistic factor for heavy atoms, $\gamma_{1/2}=\{1-(Z\cdot10^7\eta)^2\}^{1/2}$, and $a_B$ is the Bohr radius of the atom. As long as we are comparing atomic EDMs of atoms which have nuclei of similar size, the relativistic term stays the same, and the proportionality in {\bf Eq.~\ref{eq8}} will hold. While the Schiff moment scales linearly ($\propto Z$) with the atomic number, the corresponding atomic EDM scales as $\propto Z^3$, making high-Z atoms preferable for EDM searches. The $^{225}$Ra-EDM experiment at Argonne National Laboratory \cite{[19-2]} has already demonstrated the feasibility of measuring the EDM in atoms with quadrupole and octupole deformed nuclei.

\vspace{-0.25cm}
\section{RAMSEY TECHNIQUE OF SEPARATED OSCILLATING FIELDS}

A particle with non-zero MDM and EDM will show Stark and Zeeman splitting of the energy states upon the application of electric and magnetic fields. Measuring the level splitting of the energy states upon the application of electric and magnetic fields has been the standard technique for measuring the EDM of particles and atoms. The EDM is given by:
\begin{eqnarray}
d_i =\frac{\hbar\left(\omega^{B_{\uparrow},E_{\uparrow}}-\omega^{B_{\uparrow},E_{\downarrow}}\right)-2\mu_i\left(B^{B_{\uparrow},E_{\uparrow}}-B^{B_{\uparrow},E_{\downarrow}}\right)}{2\left(E^{B_{\uparrow},E_{\uparrow}}-E^{B_{\uparrow},E_{\downarrow}}\right)}\label{eq9},
\end{eqnarray}
where $\omega$ is the precession frequency, and $\mu_i$ is the MDM of the species. Here, the superscript indicates the direction of the applied magnetic ($B$) and electric fields ($E$). Usually electric and magnetic fields are characterized very precisely with the help of magnetometers. That leaves measuring the precession frequency to a high degree of precision.

The precession frequency of a polarized species is measured using the Ramsey technique of separated oscillating fields. The method involves (i) preparing a polarized ensemble of particles or atoms and subjecting them to a constant magnetic field, $B_0$, along the direction of the magnetization, (ii) flipping the ensemble magnetization by applying an oscillating magnetic field (with a frequency, $\omega_{RF}$) perpendicular to the main $B_0$ magnetic field, for a set time period, $t_{\pi/2}$, (iii) allowing the system to precess for a time period of $t_s$, (iv) applying the oscillating magnetic field, with the same frequency and duration as in step (ii), in order to flip the magnetization along an axis parallel to the magnetization of the initially prepared state. Finally the particle spins are measured and their respective counts is given by \cite{[29],[30]}:
\begin{eqnarray}
N_{\uparrow,\downarrow} = \frac{N_0}{2}\left[ 1\pm \alpha \cos \left\{ (\omega_{RF}-\gamma_i B_0) \cdot \left(t_s+\frac{4 t_{\pi/2}}{\pi}\right) \right\} \right] \label{eq10},
\end{eqnarray}
where $N_0$ is the total number of particles or atoms counted at the end, $\gamma_i$ is the gyromagnetic ratio of the species, and $\alpha$ is the initial degree of polarization. The count asymmetry between final spin up and spin down states is measured as a function of the the frequency of the applied oscillating magnetic field ($\omega_{RF}$). The global extremum of this scan \emph{w.r.t.} $\omega_{RF}$ gives an accurate measure of the precession frequency, as at the global extremum, the precession frequency is equal to the frequency of the applied oscillating magnetic field. The statistical precision for the measurement of the EDM thus obtained by determining the precession frequency using the Ramsey method is \cite{[31]}:
\begin{eqnarray}
\sigma_{d_i} \approx \frac{\hbar}{2\alpha t_s |E|\sqrt{N_0\cdot \mathcal{M}}}\label{eq11},
\end{eqnarray}
where $\mathcal{M}$ is the number of times the Ramsey cycle described above is repeated.

\vspace{-0.25cm}
\section{SURVEY OF DEFORMED NUCLEI}

In the past, though the isotopes have been studied for use in atomic EDM experiments, for the first time, we here present a comprehensive survey of the best candidates in which to measure an atomic EDM. In this survey, reported in {\bf TABLE.~\ref{tab1}}, we have attempted to (i) identifying the isotopes with the highest theoretical expectation for an EDM, according to {\bf Eq.~\ref{eq8}}, which means searching for heavy isotopes with high deformation coefficients, and a low ground state parity doublet energy difference, and also (ii) identifying those deformed isotopes which have a high rate of production at the Facility for Rare Isotope Beams (FRIB), so that we are able to achieve the best possible statistical precision in an experiment, according to {\bf Eq.~\ref{eq11}}. For the purposes of this survey, we constrained ourselves to isotopes which are at least $|j|=1/2$, and which have comparable or higher (lower) deformation parameters (ground state parity doublet energy difference) than that of $^{225}$Ra. However, we have included isotopes of Francium given that an effort to measure their EDM is underway \cite{[18-4]}. Furthermore, we restricted ourselves to long lived isotopes which have a lifetime, at least, on the order of a minute. The theoretical expectation for an EDM is dictated by {\bf Eq.~\ref{eq8}}: $d^{\mathcal{T}}\!\propto\!\{(\beta_2\beta^2_3 Z^3 A^{2/3})/(E_- - E_+)\}$.  Presented here these values have been normalized with that of $^{225}$Ra. FRIB is expected to produce large amounts of heavy nuclei that fall in the octupole deformation region on the nuclide chart. With a higher FRIB rate, an EDM experiment can achieve better relative statistical sensitivity, $\mathcal{W}\propto\sqrt{W^{(\text{FRIB})}}$ (from {\bf Eq.~\ref{eq11}}), \emph{i.e.} the experiment will be sensitive to a smaller value of EDM. Here $W^{(\text{FRIB})}$ is the stopped beam rate at FRIB from the ultimate FRIB yield in Refs. \cite{[32-1],[32-2]}. Our constraint on lifetime makes it long enough that we have not considered it as a part of the statistical sensitivity\footnote{Some isotopes may only have a lifetime of a few minutes and while some others reported here have a lifetime on the order of many thousand years. Including lifetime into the statistical precision will bias the survey to those isotopes which have a high lifetime, even though experimental spin-coherence times will restrict the maximum storage time to an order of minutes.}. Finally, what matters is the confluence of a high theoretical expectation for an EDM and the best statistical precision achievable in measurements. Towards realizing this confluence, we have reported a relative impact factor for each candidate isotope, which is $\mathcal{I}\propto \{(\beta_2\beta^2_3 Z^3 A^{2/3})/(E_- - E_+)\}\cdot\sqrt{W^{(\text{FRIB})}}$. We have not considered the issues of cooling, trapping, and spin - coherence time scales for the species in this survey. Also we have not considered the sign of the EDM. While the deformation parameters reported here come from theoretical models, such as the M\" oller-Nix model \cite{[33-2-2]}, the level scheme has not yet been well established for many isotopes. Other theoretical frameworks to compute the structure deformation parameters that were considered in this survey can be found in Refs. \cite{[33-0-1],[33-0-2]}. In {\bf TABLE.~\ref{tab2}}, we have reported those isotopes which are quadrupole and octupole deformed, but whose ground state parity doublet energy difference has not yet been measured. We have neglected reporting quadrupole and octupole deformed isotopes whose ground state is a $0^+$, since these cannot posses an EDM or an MDM, \emph{viz.} $^{224}$Ra, $^{226}$Ra, and $^{224}$Ac, to name a few.

Even for the most octupole deformed nuclei listed in {\bf TABLE.~\ref{tab1}}, the value of the deformation parameters of $\{\beta_2,\beta_3\}$ do not go higher than about $\sim\{0.21,0.16\}$, and are comparable to that of $^{225}$Ra. The enhancement to EDM mostly is governed by the ground state parity doublet energy difference, since it spans over $\sim3$ orders of magnitude. 

Octupole deformed isotopes form islands when plotted according to $\{Z \otimes (A-Z) \}$ \cite{[33-2-2],[34]}, where the value of the deformation parameters rises, peaks, and declines as one increases $Z$ (by holding $(A-Z)$ constant, and vice-versa). For example, among the Ra isotopes, the predicted island of octupole deformation occurs at $A \in (217,229)$, where the value of $\beta_3$ peaks at $A=222$. In {\bf TABLE.~\ref{tab3}}, we have listed those isotopes which do not follow this general trend. These isotopes have values of $\beta_2$ comparable to that of $^{225}$Ra, and a lifetime of at least a minute.

\begin{table}[h]
\centering
\caption{Deformed nuclei whose structure coefficients and parity doublet energy splitting are known.$^{*}$}
\label{tab1}
\begin{adjustbox}{max width=16cm}
\begin{tabular}{lrrrrrcccc}
\hline
& \multicolumn{1}{c}{$T_{1/2}$} & \multicolumn{1}{c}{$\beta_3$} & \multicolumn{1}{c}{$\beta_2$} & \multicolumn{1}{c}{$j$} & \multicolumn{1}{c}{$\Delta E~$(keV)} & $d^{\mathcal{T}}$ & $W^{(\text{FRIB})}$ & $\mathcal{W}$ & $\sim\!\mathcal{I}$\\
\hhline{==========}
$^{221}$Rn$^{\#}$ & $25(2)$ min \cite{[33-1-1]} & $0.142$ \cite{[33-2-2]} & $0.119$ \cite{[33-2-2]} & $7/2^+$ \cite{[33-1-1]} & 30(10) \cite{[33-1-2]} & $1.61$ & $2.29$ & $0.6$ & $0.97$\\
\hline
$^{221}$Fr & $4.9(2)$ min \cite{[33-1-1]} & $0.100$ \cite{[23]} & $0.106$ \cite{[23]} & $5/2^-$ \cite{[33-1-1]} & 234.51(6) \cite{[33-1-1]} & $0.09$ & $6.13$ & $0.98$ & $0.09$\\
$^{223}$Fr & $22.00(7)$ min \cite{[33-2-1]} & $0.135$ \cite{[33-2-2]} & $0.146$ \cite{[33-2-2]} & $3/2^{(-)}$ \cite{[33-2-1]} & $160.43(3)$ \cite{[33-2-1]} & $0.35$ & $4.26$ & $0.82$ & $0.29$\\
$^{227}$Fr & $2.47(3)$ min \cite{[33-4-1]} & $0.070$ \cite{[33-2-2]} & $0.181$ \cite{[33-2-2]} & $1/2^+$ \cite{[33-4-1]} & $62.97(7)$ \cite{[33-4-1]} & $0.30$ & $0.702$ & $0.33$ & $0.10$\\
\hline
$^{221}$Ra & $28(2)$ s \cite{[33-1-1]} & $0.145$ \cite{[33-2-2]} & $0.111$ \cite{[33-2-2]} & $5/2^+$ \cite{[33-1-1]} & $103.61(11)$ \cite{[33-1-1]} & $0.49$ & $6.05$ & $0.97$ & $0.48$\\
$^{223}$Ra & $11.43(5)$ day \cite{[33-2-1]} & $0.142$ \cite{[33-2-2]} & $0.156$ \cite{[33-2-2]} & $3/2^+$ \cite{[33-2-1]} & $50.128(9)$ \cite{[33-2-1]} & $1.36$ & $6.80$ & $1.03$ & $1.40$\\
$^{225}$Ra & $14.9(2)$ day \cite{[33-3-1]} & $0.124$ \cite{[33-2-2]} & $0.164$ \cite{[33-2-2]} & $1/2^+$ \cite{[33-3-1]} & $55.16(6)$ \cite{[33-3-1]} & $1.00$ & $6.41$ & $1.00$ & $1.00$\\
\hline
$^{223}$Ac & $2.10(5)$ min \cite{[33-2-1]} & $0.151$ \cite{[33-2-2]} & $0.147$ \cite{[33-2-2]} & $(5/2^-)$ \cite{[33-2-1]} & $64.62(4)$ \cite{[33-2-1]} & $1.49$ & $5.18$ & $0.90$ & $1.34$\\
$^{225}$Ac & $9.9203(3)$ day \cite{[33-3-1]} & $0.127$ \cite{[33-2-2]} & $0.164$ \cite{[33-2-2]} & $(3/2^-)$ \cite{[33-3-1]} & $40.09(5)$ \cite{[33-3-1]} & $1.49$ & $5.10$ & $0.89$ & $1.33$\\
$^{227}$Ac & $21.772(3)$ year \cite{[33-4-1]} & $0.105$ \cite{[33-2-2]} & $0.172$ \cite{[33-4-1]} & $3/2^-$ \cite{[33-4-1]} & $27.369(11)$ \cite{[33-4-1]} & $1.56$ & $6.40$ & $1.00$ & $1.56$\\
\hline
$^{229}$Th & $7880(120)$ year \cite{[33-5-1]} & $0.240$ \cite{[33-5-2]} & $0.115$ \cite{[33-5-2]} & $5/2^+$ \cite{[33-5-1]} & $146.3569(14)$ \cite{[33-5-1]} & $1.07$ & $6.19$ & $0.98$ & $1.05$\\
\hline
$^{229}$Pa & $1.50(5)$ day \cite{[33-5-1]} & $0.082$ \cite{[23]} & $0.190$ \cite{[33-2-2]} & $5/2^+$ \cite{[33-6-1]} & $0.06(5)$ \cite{[33-6-1]} & $521.13$ & $20.7$ & $1.8$ & $938.03$\\
\hhline{==========}
\end{tabular}
\end{adjustbox}
\tablenote[t1n1]{The theoretical expectation of EDM ($d^{\mathcal{T}}$, {\bf Eq.~\ref{eq8}}), relative statistical sensitivity ($\mathcal{W}=\sqrt{W^{(\text{FRIB})}}$, {\bf Eq.~\ref{eq11}}), and the relative impact ($\mathcal{I}$) have all been normalized to $^{225}$Ra. Units of beam rate, $W^{(\text{FRIB})}$ is $10^6/$s. $\Delta E = E_--E_+$. \\ $^{\#}$ Parity doublet energy levels have not been adopted by the National Nuclear Data Center (NNDC) database.}
\end{table}

\begin{table}[h]
\centering
\caption{Deformed nuclei whose parity doublet energy splitting are not known.}
\label{tab2}
\begin{adjustbox}{max width=12cm}
\begin{tabular}{lrrrrcc}
\hline
& \multicolumn{1}{c}{$T_{1/2}$} & \multicolumn{1}{c}{$\beta_3$} & \multicolumn{1}{c}{$\beta_2$} & \multicolumn{1}{c}{$j$} & $W^{(\text{FRIB})}$ & $\mathcal{W}$\\
\hhline{=======}
$^{223}$Rn & $24.3(4)$ min \cite{[33-2-1]} & $0.117$ \cite{[33-2-2]} & $0.155$ \cite{[33-2-2]} & $7/2^{(+)}$ \cite{[33-2-1]} & $0.903$ & $0.38$ \\
$^{225}$Rn & $4.66(4)$ min \cite{[33-3-1]} & $0.077$ \cite{[33-2-2]} & $0.163$ \cite{[33-2-2]} & $7/2^-$ \cite{[33-3-1]} & $0.271$ & $0.21$ \\
\hline
$^{225}$Fr$^{*}$ & $3.95(14)$ min \cite{[33-3-1]} & $0.108$ \cite{[33-2-2]} & $0.163$ \cite{[33-2-2]} & $3/2^-$ \cite{[33-3-1]} & $2.08$ & $0.57$\\
\hline
$^{226}$Ac$^{\#}$ & $29.37(12)$ hour \cite{[33-7-1]} & $0.120$ \cite{[33-2-2]} & $0.164$ \cite{[33-2-2]} & $1^{(+)}$ \cite{[33-7-1]} & $5.78$ & $0.95$\\
\hhline{=======}
\end{tabular}
\end{adjustbox}
\tablenote[t2n1]{No parity doublets have been observed, \cite{[33-7-2-1]}; $^{\#}$ \cite{[33-7-2-2]}.}
\end{table}
\begin{table}[h]
\centering
\caption{Suspect deformed nuclei.}
\label{tab3}
\begin{adjustbox}{max width=12cm}
\begin{tabular}{lrrlrcc}
\hline
& \multicolumn{1}{c}{$T_{1/2}$} & \multicolumn{1}{c}{$\beta_2$} & \multicolumn{1}{c}{$j$} & \multicolumn{1}{c}{$\Delta E$} & $W^{(\text{FRIB})}$ & $\mathcal{W}$\\
\hhline{=======}
$^{231}$Ac & $7.5(1)$ min \cite{[33-8-1]} & $0.207$ \cite{[33-2-2]} & $1/2^+$  \cite{[33-8-1]} & $372.28(8)$ \cite{[33-8-1]} & $5.27$ & $0.91$\\
\hline
$^{227}$Th & $18.697(7)$ day \cite{[33-4-1]} & $0.173$ \cite{[33-2-2]} & $(1/2^+)$ \cite{[33-4-1]} & $67.2(2)$ \cite{[33-9-5]} & $8.91$ & $1.18$\\
\hline
$^{226}$Pa & $1.8(2)$ min \cite{[33-7-1]} & $0.165$ \cite{[33-2-2]} & $1^{+\,*}$&  \multicolumn{1}{c}{$^{?}$} & $19.7$ & $1.75$\\
\hhline{=======}
\end{tabular}
\end{adjustbox}
\tablenote[t3n1]{Inferred from $\alpha$-decay to $^{222}$Ac \cite{[33-9-4]}. $^{?}$ Indicates that the parity doublet has not yet been observed.}
\end{table}

\vspace{-0.25cm}
\section{CONCLUSION}

Although the theoretical models predict the magnetic moment and ground state spin of the nuclei well \cite{[23],[35],[36],[37]}, they mutually disagree on the values of nuclear deformation parameters. The values of deformation parameters sometimes vary by as much as $\sim50\%$ between various theoretical models. For example, the value of $\beta_3$ for $^{225}$Ra is $0.124$ in the Moller-Nix model \cite{[33-2-2]}; $0.146$ in the Hartree-Fock model \cite{[22]}; and $0.104$ using Nilsson model \cite{[22]}. Variations in EDM enhancement ($d^{\mathcal{T}},~\mathcal{I}$) below a factor of $\sim2$ are within the spread of deformation parameters from the theoretical models. The only isotope in {\bf TABLE.~\ref{tab1}} which has a greatly enhanced atomic EDM compared to $^{225}$Ra is $^{229}$Pa. This enhancement comes mostly from the smallness of the energy difference for its ground state parity doublet, since the deformation parameters are comparable to that of $^{225}$Ra. It is important to note that the recent measurement of its parity doublet energy difference, $0.06(5)~$keV \cite{[33-6-1]}, is associated with a large relative uncertainty so that the authors indicate that the existence of the doublet is uncertain.
An earlier measurement by the same group had proposed existence of the doublet with an energy difference of $0.22(5)~$keV \cite{[33-6-1-1]} which would result in a factor of 4 reduction in $\mathcal{I}$.
It may be interesting to measure the ground state parity doublet energy difference for isotopes in {\bf TABLE.~\ref{tab2}}, and also re-calculate the deformation parameters using an independent theoretical framework than the ones from which the deformation parameter values have been reported here for isotopes in {\bf TABLE.~\ref{tab3}}. We conclude that $^{225}$Ra is one of the best systems in which to measure an atomic EDM, since its theoretical EDM expectation is among the most enhanced, and $^{229}$Pa may provide for another system in which to attempt a measurement of the atomic EDM.

\vspace{-0.25cm}
\section{ACKNOWLEDGMENTS}

The authors would like to thank Dr. Matthew Dietrich and Prof. Dr. Jonathan Engel for useful discussions. The authors would particularly like to acknowledge the support provided by Prof. Dr. Anatoli Afanasjev's group towards realizing the theoretical calculations of the nuclear deformation parameters and the ground state parity doublet energy difference for the isotopes highlighted in this paper. One of the authors (P.~M.) is supported by SERI-FCS award \# 2015.0594 and Sigma Xi grant \# G2017100190747806. The remaining authors are supported by DOE grant \# DE-SC0014448. The authors would also like to acknowledge the travel support provided by their respective departments and institutes.


\nocite{*}
\vspace{-0.25cm}
\bibliographystyle{aipnum-cp}%
\renewcommand*{\bibfont}{\small}

\begin{thebibliography}{16cm}
\bibitem{[3-1]}{W. Pauli, Il Nuovo Cimento {\bf 6}, 204 (1957). DOI: \href{https://doi.org/10.1007/BF02827771}{10.1007/BF02827771}.}
\bibitem{[3-2]}{G. L\" uders, Ann. Phys. {\bf 2}, 1 (1957). DOI: \href{https://doi.org/10.1016/0003-4916(57)90032-5}{10.1016/0003-4916(57)90032-5}.}
\bibitem{[3-3]}{O. W. Greenberg, Phys. Rev. Lett. {\bf 89}, 231602 (2002). DOI: \href{http://dx.doi.org/10.1103/PhysRevLett.89.231602}{10.1103/PhysRevLett.89.231602}. 
}

\bibitem{[1]}{G. Steigman, Annu. Rev. Nucl. Part. Sci. {\bf 57}, 463 (2007). DOI: \href{https://doi.org/10.1146/annurev.nucl.56.080805.140437}{10.1146/annurev.nucl.56.080805.140437}. 
}
\bibitem{[2]}{N. Aghanim et. al. (Planck Collaboration), (2018). arXiv: \href{http://arxiv.org/abs/1807.06209}{[1807.06209]}.
}
\bibitem{[3]}{A. D. Sakharov, JETP Lett. {\bf 5}, 1 (1967): pp. 24-27. DOI: \href{http://dx.doi.org/10.1070/PU1991v034n05ABEH002497}{10.1070/PU1991v034n05ABEH002497}.}

\bibitem{[5-1]}{K. Abe et. al. (Super-K Collaboration), Phys. Rev. D {\bf 91}, 072006 (2015). DOI: \href{http://dx.doi.org/10.1103/PhysRevD.91.072006}{10.1103/PhysRevD.91.072006}. 
}
\bibitem{[5-2-1]}{A. P. Serebrov et. al., Nucl. Instrum. Methods A {\bf 137} (2008). DOI: \href{http://dx.doi.org/10.1016/j.nima.2009.07.041}{10.1016/j.nima.2009.07.041}. 
}
\bibitem{[5-2-2]}{Z. Berezhiani et. al., Eur. Phys. J. C {\bf 78}, 717 (2018). DOI: \href{http://dx.doi.org/10.1140/epjc/s10052-018-6189-y}{10.1140/epjc/s10052-018-6189-y}. 
}
\bibitem{[5-2-3]}{C. Abel et. al. (PSI-nEDM Collaboration), (2018). arXiv: \href{http://arxiv.org/abs/1811.01906}{[1811.01906]}.}
\bibitem{[5-3]}{G. Gabrielse et. al., Phys. Rev. Lett. {\bf 65}, 1317 (1990). DOI: \href{http://dx.doi.org/10.1103/PhysRevLett.65.1317}{10.1103/PhysRevLett.65.1317}.}
\bibitem{[5-4]}{G. R. Farrar and M. E. Shaposhnikov, Phys. Rev. Lett. {\bf 70}, 2833 (1993). DOI: \href{http://dx.doi.org/10.1103/PhysRevLett.70.2833}{10.1103/PhysRevLett.70.2833}. 
}
\bibitem{[5-5]}{A. Riotto and M. Trodden, Annu. Rev. Nucl. Part. Sci. {\bf 49}, 35 (1999). DOI: \href{https://doi.org/10.1146/annurev.nucl.49.1.35}{10.1146/annurev.nucl.49.1.35}. 
}

\bibitem{[4]}{J. H. Christenson et. al., Phys. Rev. Lett. {\bf 13}, 138 (1964). DOI: \href{https://doi.org/10.1103/PhysRevLett.13.138}{10.1103/PhysRevLett.13.138}.}
\bibitem{[5]}{B. Aubert et. al. (BABAR Collaboration), Phys. Rev. Lett. {\bf 89}, 201802 (2002). DOI: \href{https://doi.org/10.1103/PhysRevLett.89.201802}{10.1103/PhysRevLett.89.201802}. 
}

\bibitem{[5-6]}{G. ’t Hooft, Phys. Rev. Lett. {\bf 37}, 8 (1976). DOI: \href{https://link.aps.org/doi/10.1103/PhysRevLett.37.8}{10.1103/PhysRevLett.37.8}.}
\bibitem{[5-7]}{J. Ellis, Nucl. Instrum. Methods A {\bf 284}, 33 (1989). DOI: \href{http://dx.doi.org/10.1016/0168-9002(89)90243-X}{10.1016/0168-9002(89)90243-X}.}

\bibitem{[14]}{T. E. Chupp et. al., Rev. Mod. Phys. {\bf 91}, 015001 (2019). DOI: \href{http://dx.doi.org/10.1103/RevModPhys.91.015001}{10.1103/RevModPhys.91.015001}. 
}
\bibitem{[15]}{D. Geesaman, Nuclear Physics News {\bf 26}, 3 (2016). DOI: \href{http://dx.doi.org/10.1080/10619127.2016.1140967}{10.1080/10619127.2016.1140967}.}

\bibitem{[6-0-1]}{E. M. Purcell and N. F. Ramsey, Phys. Rev. {\bf 78}, 807 (1950). DOI: \href{http://dx.doi.org/10.1103/PhysRev.78.807}{10.1103/PhysRev.78.807}.}
\bibitem{[16]}{L. I. Schiff, Phys. Rev. {\bf 132}, 2194 (1963). DOI: \href{http://dx.doi.org/10.1103/PhysRev.132.2194}{10.1103/PhysRev.132.2194}.}
\bibitem{[23]}{V. Spevak et. al., Phys. Rev. C {\bf 56}, 1357 (1997). DOI: \href{http://dx.doi.org/10.1103/PhysRevC.56.1357}{10.1103/PhysRevC.56.1357}. 
}
\bibitem{[6-0-2]}{T. Fukuyama, Int. J. Mod. Phys. A {\bf 27}, 1230015 (2012). DOI: \href{http://dx.doi.org/10.1142/S0217751X12300153}{10.1142/S0217751X12300153}. 
}

\bibitem{[6]}{K. T. Hecht, Quantum Mechanics (Springer, NY), (2002): pp. 299-302. DOI: \href{https://doi.org/10.1007/978-1-4612-1272-0_32}{10.1007/978-1-4612-1272-0\_32}.}

\bibitem{[7]}{W. Bernreuther and M. Suzuki, Rev. Mod. Phys. {\bf 63}, 313 (1991). DOI: \href{http://dx.doi.org/10.1103/RevModPhys.63.313}{10.1103/RevModPhys.63.313}.}

\bibitem{[8]}{I. B. Khriplovich and A. R. Zhitnitsky, Phys. Lett. B {\bf 109}, 490 (1982). DOI: \href{http://dx.doi.org/10.1016/0370-2693(82)91121-2}{10.1016/0370-2693(82)91121-2}.}
\bibitem{[9]}{M. Pospelov and A. Ritz, Ann. Phys. {\bf 318} 119 (2005). DOI: \href{http://dx.doi.org/10.1016/j.aop.2005.04.002}{10.1016/j.aop.2005.04.002}. 
}
\bibitem{[10]}{J. M. Pendlebury et. al., Phys. Rev. D {\bf 92}, 092003 (2015). DOI: \href{http://dx.doi.org/10.1103/PhysRevD.92.092003}{10.1103/PhysRevD.92.092003}. 
}
\bibitem{[11]}{M. Pospelov and A. Ritz, Phys. Rev. D {\bf 89}, 056006 (2014). DOI: \href{http://dx.doi.org/10.1103/PhysRevD.89.056006}{10.1103/PhysRevD.89.056006}. 
}
\bibitem{[12]}{D. Ghosh and R. Sato, Phys. Lett. B {\bf 777}, 335 (2018). DOI: \href{http://dx.doi.org/10.1016/j.physletb.2017.12.052}{10.1016/j.physletb.2017.12.052}. 
}
\bibitem{[12-2]}{J. Baron et. al. (ACME Collaboration), Science {\bf 343}, 269 (2014). DOI: \href{http://dx.doi.org/10.1126/science.1248213}{10.1126/science.1248213}. 
}

\bibitem{[17]}{C.-P. Liu et. al., Phys. Rev. C {\bf 76}, 035503 (2007). DOI: \href{http://dx.doi.org/10.1103/PhysRevC.76.035503}{10.1103/PhysRevC.76.035503}. 
}

\bibitem{[18-1]}{E. S. Ensberg, Phys. Rev. {\bf 153}, 36 (1967). DOI: \href{http://dx.doi.org/10.1103/PhysRev.153.36}{10.1103/PhysRev.153.36}.}
\bibitem{[18-2]}{S. A. Murthy et. al., Phys. Rev. Lett. {\bf 63}, 965 (1989). DOI: \href{http://dx.doi.org/10.1103/PhysRevLett.63.965}{10.1103/PhysRevLett.63.965}.}
\bibitem{[18-3]}{E. D. Commins et. al., Phys. Rev. A {\bf 50}, 2960 (1994). DOI: \href{http://dx.doi.org/10.1103/PhysRevA.50.29605}{10.1103/PhysRevA.50.2960}.}
\bibitem{[18-4]}{T. Inoue et. al., Hyperfine Interact. {\bf 231}, 157 (2015). DOI: \href{https://doi.org/10.1007/s10751-014-1100-1}{10.1007/s10751-014-1100-1}.}

\bibitem{[19-1]}{E. R. Tardiff et. al., Hyperfine Interact. {\bf 225}, 197 (2014). DOI: \href{https://doi.org/10.1007/s10751-013-0898-2}{10.1007/s10751-013-0898-2}.}
\bibitem{[19-2]}{M. Bishof et. al., Phys. Rev. C {\bf 94}, 025501 (2016). DOI: \href{http://dx.doi.org/10.1103/PhysRevC.94.025501}{10.1103/PhysRevC.94.025501}. 
}

\bibitem{[20-1]}{F. Allmendinger et. al., Phys. Rev. A {\bf 100}, 022505 (2019). DOI: \href{https://doi.org/10.1103/PhysRevA.100.022505}{10.1103/PhysRevA.100.022505}.}
\bibitem{[20-2]}{B. Graner et. al., Phys. Rev. Lett. {\bf 116}, 161601 (2016). DOI: \href{http://dx.doi.org/10.1103/PhysRevLett.116.161601}{10.1103/PhysRevLett.116.161601}. 
}

\bibitem{[21]}{J. Engel et. al., Prog. Part. Nucl. Phys. {\bf 71}, 21 (2013). DOI: \href{http://dx.doi.org/10.1016/j.ppnp.2013.03.003}{10.1016/j.ppnp.2013.03.003}. 
}

\bibitem{[22]}{J. Engel et. al., Phys. Rev. C {\bf 68}, 025501 (2003). DOI: \href{http://dx.doi.org/10.1103/PhysRevC.68.025501}{10.1103/PhysRevC.68.025501}. 
}

\bibitem{[24]}{N. Auerbach et. al., Phys. Rev. Lett. {\bf 76}, 4316 (1996). DOI: \href{http://dx.doi.org/10.1103/PhysRevLett.76.4316}{10.1103/PhysRevLett.76.4316}. 
}

\bibitem{[25-1]}{W. D. Myers and W. J. Swiatecki, Ann. Phys. {\bf 55}, 395 (1969). DOI: \href{http://dx.doi.org/10.1016/0003-4916(69)90202-4}{10.1016/0003-4916(69)90202-4}.}
\bibitem{[25-2]}{G. A. Leander et. al., Nucl. Phys. A {\bf 453}, 58 (1986). DOI: \href{http://dx.doi.org/10.1016/0375-9474(86)90029-1}{10.1016/0375-9474(86)90029-1}.}
\bibitem{[25-3]}{W. Nazarewicz, Nucl. Phys. A {\bf 520}, c333 (1990). DOI: \href{http://dx.doi.org/10.1016/0375-9474(91)90489-S}{10.1016/0375-9474(91)90489-S}.}
\bibitem{[25-4]}{P. A. Butler and W. Nazarewicz, Nucl. Phys. A {\bf 533}, 249 (1991). DOI: \href{http://dx.doi.org/10.1016/0375-9474(91)90489-S}{10.1016/0375-9474(91)90489-S}.}

\bibitem{[26]}{V. A. Dzuba et. al., Phys. Rev. A {\bf 66}, 012111 (2002). DOI: \href{http://dx.doi.org/10.1103/PhysRevA.66.012111}{10.1103/PhysRevA.66.012111}. 
}
\bibitem{[27]}{V. V. Flambaum, Phys. Rev. C {\bf 99}, 035501 (2019). DOI: \href{http://dx.doi.org/10.1103/PhysRevC.99.035501}{10.1103/PhysRevC.99.035501}. 
}
\bibitem{[28]}{O. P. Sushkov et. al., JETP Letters {\bf 60}, 5 (1984): pp. 873. URL: \href{http://www.jetp.ac.ru/cgi-bin/e/index/e/60/5/p873?a=list}{www.jetp.ac.ru/cgi-bin/e/index/e/60/5/p873?a=list}.}

\bibitem{[29]}{N. F. Ramsey, Phys. Rev. {\bf 76}, 996 (1949). DOI: \href{http://dx.doi.org/10.1103/PhysRev.76.996}{10.1103/PhysRev.76.996}.}
\bibitem{[30]}{N. F. Ramsey, Phys. Rev. {\bf 78}, 695 (1950). DOI: \href{http://dx.doi.org/10.1103/PhysRev.78.695}{10.1103/PhysRev.78.695}.}
\bibitem{[31]}{N. F. Ramsey, Physica B+C {\bf 137}, 223 (1986). DOI: \href{http://dx.doi.org/10.1016/0378-4363(86)90326-8}{10.1016/0378-4363(86)90326-8}.}
\bibitem{[32-1]}{K. S\" ummerer and B. Blank, Phys. Rev. C {\bf 61}, 034607 (2000). DOI: \href{http://dx.doi.org/10.1103/PhysRevC.61.034607}{10.1103/PhysRevC.61.034607}.}
\bibitem{[32-2]}{O. B. Tarasov and D. Bazin, Nucl. Instrum. Methods B {\bf 266}, 4657 (2008). DOI: \href{http://dx.doi.org/10.1016/j.nimb.2008.05.110}{10.1016/j.nimb.2008.05.110}.}
\bibitem{[33-0-1]}{A. V. Afanasjev et. al., Phys. Scr. {\bf 93}, 034002 (2018). DOI: \href{http://dx.doi.org/10.1103/PhysRevC.93.044304}{10.1103/PhysRevC.93.044304}.}
\bibitem{[33-0-2]}{S. E. Agbemava et. al., Phys. Rev. C {\bf 93}, 044304 (2016). DOI: \href{http://dx.doi.org/10.1016/j.nimb.2008.05.110}{10.1016/j.nimb.2008.05.110}.}
\bibitem{[33-1-1]}{A. Kumar Jain et. al., Nucl. Data Sheets {\bf 108}, 883 (2007). DOI: \href{http://dx.doi.org/10.1016/j.nds.2007.03.002}{10.1016/j.nds.2007.03.002}.}
\bibitem{[33-2-2]}{P. Moller et. al., At. Data Nucl. Data Tables {\bf 59}, 185 (1995). DOI: \href{http://dx.doi.org/10.1006/adnd.1995.1002}{10.1006/adnd.1995.1002}.}
\bibitem{[33-1-2]}{C. F. Liang et. al., Phys. Rev. C {\bf 62}, 047303 (2000). DOI: \href{http://dx.doi.org/10.1103/PhysRevC.62.047303}{10.1103/PhysRevC.62.047303}.}
\bibitem{[33-2-1]}{E. Browne, Nucl. Data Sheets {\bf 93}, 763 (2001). DOI: \href{http://dx.doi.org/10.1006/ndsh.2001.0016}{10.1006/ndsh.2001.0016}.}
\bibitem{[33-4-1]}{F. Kondev et. al., Nucl. Data Sheets {\bf 132}, 257 (2016). DOI: \href{http://dx.doi.org/10.1016/j.nds.2016.01.002}{10.1016/j.nds.2016.01.002}.}
\bibitem{[33-3-1]}{A. K. Jain et. al.,, Nucl. Data Sheets {\bf 110}, 1409 (2009). DOI: \href{http://dx.doi.org/10.1006/adnd.1995.1002}{10.1006/adnd.1995.1002}.}
\bibitem{[33-5-1]}{E. Browne and J. K. Tuli, Nucl. Data Sheets {\bf 109}, 2657 (2008). DOI: \href{http://dx.doi.org/10.1016/j.nds.2008.10.001}{10.1016/j.nds.2008.10.001}.}
\bibitem{[33-5-2]}{N. Minkov and A. P\'alffy, Phys. Rev. Lett. {\bf 118}, 212501 (2017). DOI: \href{http://dx.doi.org/10.1103/PhysRevLett.118.212501}{10.1103/PhysRevLett.118.212501}.}
\bibitem{[33-6-1]}{I. Ahmad et. al., Phys. Rev. C {\bf 92}, 024313 (2015). DOI: \href{http://dx.doi.org/10.1103/PhysRevC.92.024313}{10.1103/PhysRevC.92.024313}.}
\bibitem{[33-7-1]}{Y. A. Akovali, Nucl. Data Sheets {\bf 77}, 433 (1996). DOI: \href{http://dx.doi.org/10.1006/ndsh.1996.0005}{10.1006/ndsh.1996.0005}.}
\bibitem{[33-7-2-1]}{D. G. Burke et. al., Nucl. Phys. A {\bf 612}, 91 (1997). DOI: \href{http://dx.doi.org/10.1016/S0375-9474(96)00311-9}{10.1016/S0375-9474(96)00311-9}.}
\bibitem{[33-7-2-2]}{J. D. McCoy, Soc. Sci. Fennica, Commentationes Phys. -Math. {\bf 30}, (1964). DOI: \href{http://dx.doi.org/10.1016/S0375-9474(96)00311-9}{10.1016/S0375-9474(96)00311-9}.}
\bibitem{[33-8-1]}{E. Browne and J. K. Tuli, Nucl. Data Sheets {\bf 114}, 751 (2013). DOI: \href{http://dx.doi.org/10.1016/j.nds.2013.05.002}{10.1016/j.nds.2013.05.002}.}
\bibitem{[33-9-5]}{C. F. Liang et. al., Phys. Rev. C {\bf 51}, 1199 (1995). DOI: \href{http://dx.doi.org/10.1103/physrevc.51.1199}{10.1103/physrevc.51.1199}.}
\bibitem{[33-9-4]}{S. Singh et. al., Nucl. Data Sheets {\bf 112}, 2851 (2011). DOI: \href{http://dx.doi.org/10.1016/j.nds.2011.10.002}{10.1016/j.nds.2011.10.002}.}
\bibitem{[34]}{S. Ebata and T. Nakatsukasa, Phys. Scr. {\bf 92}, 064005 (2017). DOI: \href{http://dx.doi.org/10.1088/1402-4896/aa6c4c}{10.1088/1402-4896/aa6c4c}.}
\bibitem{[35]}{I. Ragnarsson, Phys. Lett. B {\bf 130}, 353 (1983). DOI: \href{http://dx.doi.org/10.1016/0370-2693(83)91520-4}{10.1016/0370-2693(83)91520-4}.}
\bibitem{[36]}{G. A. Leander and R. K. Sheline, Nucl. Phys. A {\bf 413}, 375 (1984). DOI: \href{http://dx.doi.org/10.1016/0375-9474(84)90417-2}{10.1016/0375-9474(84)90417-2}.}
\bibitem{[37]}{G. A. Leander and Y. S. Chen, Phys. Rev. C {\bf 37}, 2744 (1988). DOI: \href{http://dx.doi.org/10.1103/PhysRevC.37.2744}{10.1103/PhysRevC.37.2744}.}
\bibitem{[33-6-1-1]}{I. Ahmad et. al., Phys. Rev. Lett. {\bf 49}, 1758 (1982). DOI: \href{http://dx.doi.org/10.1103/PhysRevLett.49.1758}{10.1103/PhysRevLett.49.1758}.}
\end{thebibliography}

\end{document}